# Junction-Less Monolayer MoS$_2$ FETs

Wei Cao, Jiahao Kang, and Kaustav Banerjee

Department of Electrical and Computer Engineering, University of California, Santa Barbara, CA 93106 USA
(E-mails: weicao@ece.ucsb.edu; jiahao_kang@ece.ucsb.edu; kaustav@ece.ucsb.edu)

**Abstract**

This paper introduces monolayer molybdenum disulfide (MoS$_2$) based junction-less (JL) field-effect transistor (FET) and evaluates its performance at the smallest foreseeable (5.9 nm) transistor channel length as per the International Technology Roadmap for Semiconductors (ITRS), by employing rigorous quantum transport simulations. By comparing with MoS$_2$ based conventional FETs, it is found that the JL structure naturally lends MoS$_2$ FETs with superior device electrostatics, and higher ON-current for both high-performance and low-standby-power applications, especially at high impurity doping densities. Along with the advantages of the MoS$_2$ JL-FETs, the effects of impurity scattering induced carrier mobility degradation of JL-FETs is also highlighted as a key technological issue to be addressed for exploiting their unique features.

*Key words*— **MoS$_2$, transition metal dichalcogenides, 2D semiconductor, Junction-less FET.**

## Introduction

Junction-less FETs are considered promising candidates for next generation very-large-scale-integrated circuit (VLSI) applications because the required steep doping profile at the source/drain junctions in conventional (Conv.) nanoscale MOSFETs, which is one of the most challenging processes in complementary-metal-oxide-semiconductor (CMOS) technology, can be avoided. In other words, JL-FETs can outperform Conv. FETs (as long as device performance is not degraded) in terms of processing complexity and resultant performance variations and cost [1],[2]. In a proper JL-FET design, according to previous studies, a heavily doped and ultra-thin channel should be used. Heavy doping is a necessity to achieve ohmic source/drain contacts. Small channel thickness is required to enable full depletion of the heavily doped channel in OFF-state, and thereby suppress subthreshold leakage current [1]. In this regard, the emerging 2-dimensional (2D) or monolayer transition-metal dichalcogenide (TMD) semiconductors [3]-[5],[7]-[10], such as MoS$_2$, seem to be attractive choices because of their atomic scale thickness. Note that previously reported 2D FETs that rely on unintentional doping [5] (generally low and uncontrollable), or those fabricated with bottom gated structure [10] (non-scalable, only for academic study) can't be considered as 2D JL-FETs. On the other hand, doping techniques for monolayer MoS$_2$ as well as other 2D semiconductors are still far from being as effective and mature as that for the widely studied bulk materials, i.e., realizing steep doping profile in these 2D materials is even more challenging. In this context, JL structure may be a more natural and practical choice, compared to Conv. structures, to exploit these 2D materials in FET application. Therefore, a predictive and quantitative performance evaluation, at this stage, of the emerging 2D semiconductors based JL-FETs would be insightful to the development of both JL-FET and 2D FET technologies. In this letter, rigorous quantum simulations are employed to compare the critical metrics of the proposed monolayer MoS$_2$ (the most widely studied 2D semiconductor) based JL-FET and Conv. FET, based on which the electrostatic advantage of the JL-FET structure is found to be promising in low-standby-power applications, while the possible mobility degradation issue turns out to be a challenge for this device in high-performance applications. It is also found that the ultra-thin 2D materials, compared to bulk materials, allow higher volume doping level in JL-FET, which is desired for contact engineering. The proposal and findings provide an attractive pathway for the CMOS community that is at the verge of transferring from the 50-year-old bulk technology to the 2D materials arena.

## Device Structures and Simulation Methodology

Fig. 1(a) and (b) schematically show the device structures and doping profiles of Conv. and JL FETs, respectively. A double-gated FET topology is adopted in the simulation, in which $HfO_2$ (dielectric constant $\varepsilon_{HfO2}$ = 23.5) is used as the gate oxide. All other parameters are set according to ITRS data for 5.9-nm node [6], and are listed in Table I. The only difference of JL-FETs from Conv. FETs is that impurity doping density in JL-FET's channel is the same as that in their source/drain regions. The doping profile at the source/drain junctions of Conv. FETs normally has a gradient (see circles in the bottom plot of Fig. 1(a)) due to dopant diffusion. The nanometer-scale dimensions of modern FETs require suppressing this diffusion and thereby achieving an ultra-sharp doping gradient, which imposes severe limitations on the process thermal budget and necessitates the development of novel annealing techniques. To evaluate the intrinsic performance of Conv. FETs in this simulation work, the source/drain junctions are assumed to be ideally abrupt for simplicity. Source/drain contacts are assumed to be ohmic in this work, which can be achieved by proper contact engineering [8].

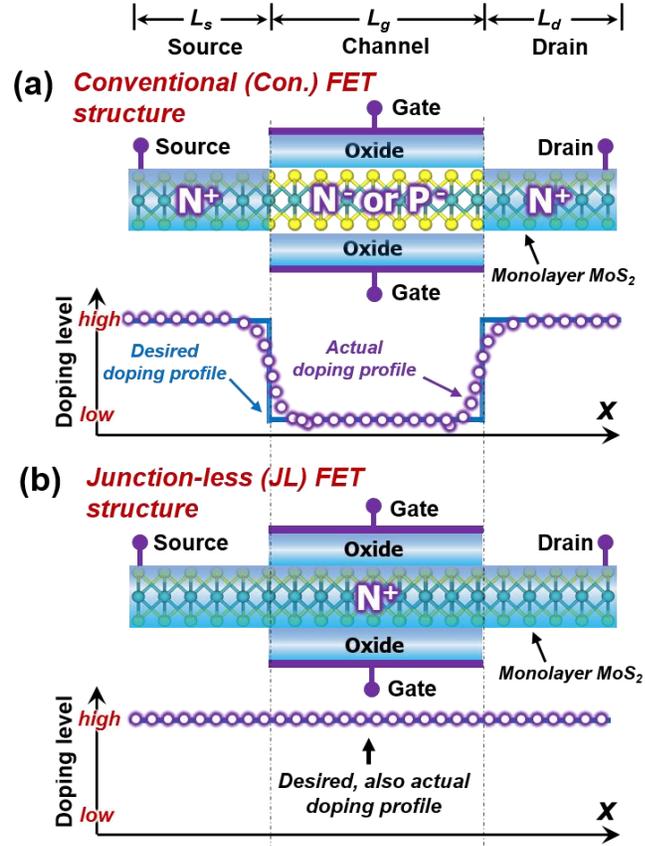

Fig. 1. Schematic view of a monolayer $MoS_2$ based double-gated (a) Conv. FET and its doping profile, and (b) JL FET and its doping profile along the channel. In Conv. FETs, it is very challenging to achieve the desired abrupt doping profile, while there is no such problem in JL FETs at all.

Device simulation is carried out with an in-house 2D FET quantum simulator in which Poisson' equation and non-equilibrium Green's function (NEGF) are self-consistently solved [4] Scattering events are properly treated through the approach of "Büttiker probes". The band structures of channel materials are properly treated during establishing the effective mass Hamiltonian. For $MoS_2$, the lowest and second lowest valleys are considered, while for Si, the lowest three quantized bands are considered, the validity of which has been systematically verified [11]. Detailed information about the simulator, as well as the approach of "Büttiker probes" can be found in [11],[12].

Table I. Parameters used in the simulation.

|      | $L_{g/s/d}$ (nm) | $T_{HfO2}$ (nm) | $V_{dd}$ (V) | $I_{off}$ ($\mu A/\mu m$) |
|------|------------------|-----------------|--------------|---------------------------|
| HP   | 5.9              | 2.7             | 0.57         | $10^{-1}$                 |
| LSTP | 5.9              | 2.7             | 0.54         | $10^{-5}$                 |

**$L_{g/s/d}$ is the length of gate/source/drain extension, $T_{HfO2}$ is the thickness of $HfO_2$ top/bottom gate dielectrics, and $V_{dd}$ is the supply voltage. $I_{off}$ is the OFF-current. HP and LSTP are abbreviations of high-performance and low-standby-power technologies, respectively.

## Results and Discussion

Fig. 2(a) shows the transfer characteristics of $MoS_2$ based JL-FET and Conv. FET, as well as that of Si based JL-FET and Conv. FET for comparison. The thickness of Si film is conservatively set to be 2 nm, which is much smaller than that projected by ITRS up to 2026 [6]. The impurity doping density in the channel or source/drain extension is set to be $5\times10^{20}$ cm$^{-3}$. It can be observed that $MoS_2$ devices have much steeper sub-threshold slopes compared to Si devices, primarily due to their much smaller thicknesses (0.65 nm for monolayer $MoS_2$) leading to stronger gate controllability and higher immunity to short-channel effects. Moreover, both JL-FETs show improved sub-threshold performance compared to Conv. FETs, which is in agreement with previous studies of Si JL-FETs. The improvement is related to the increased depletion lengths in the source/drain regions, which lead to increased "effective electrostatic channel length" [2] as reflected by the wider potential barrier in the conduction band diagrams in Fig. 2(b) for $MoS_2$ based JL-FET compared to that for $MoS_2$ based Conv. FET. In this comparison, the doping density (in relevant regions) and current level (in sub-threshold regime) are made the same for all devices. This increased effective channel length improves the device electrostatics and thus suppresses the short-channel effect. Figs. 2(c) and (d) show the relation between SS and doping density for Si and $MoS_2$ FETs, respectively. Here the lowest doping density is restricted to a relatively high value $\sim 5\times10^{19}$cm$^{-3}$, which is necessary to alleviate possible source/drain contact issue and achieve ohmic contacts [9]. Note that although high doping of 2D materials is challenging, great progress have been recently made by 2D material and device researchers that lend sufficient credibility to achieving highly doped 2D semiconductors in the near future [9],[10]. Two main effects can be found from Figs. 2(c) and (d). One is that the SS is degraded with higher doping density for all FETs under study, the other is that the SS improvement of JL-FETs, i.e., the reduction in the SS value of JL FET w.r.t that of the Conv. FET, becomes larger with higher doping density. These can be explained by the fact that higher the doping density, the shorter the effective channel length of FETs, and hence the larger the SS, which implies that in FETs with worse electrostatics, the advantage of using the JL-FET structure becomes more prominent.

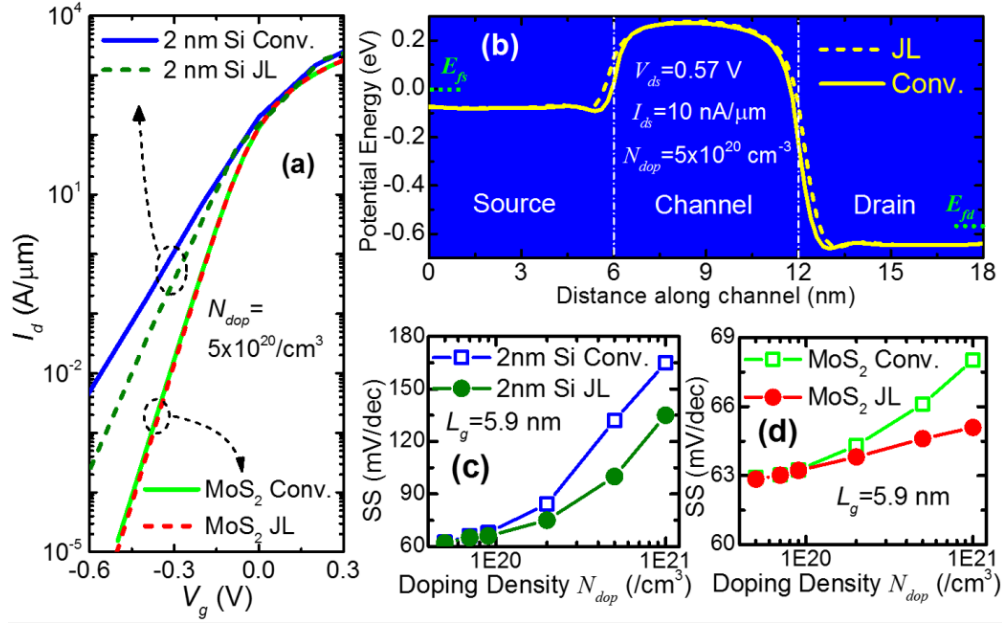

Fig. 2. (a) Transfer characteristics of 2-nm Si and monolayer $MoS_2$ based Conv. and JL-FETs with the same doping density. Here the gate metal work-functions of JL devices are tuned to shift their threshold voltages close to that of Conv. devices. (b) Conduction band diagrams for $MoS_2$ based JL and Conv. FETs with the same doping density and current levels. (c) SS versus doping density for (c) Si devices and (d) $MoS_2$ devices with both Conv. and JL-FET structures.

Although lower doping density is beneficial for obtaining better device electrostatics and hence lower SS, it introduces higher series resistance in the source/drain extensions that are not under gate control. Therefore, the doping density should be optimized, in order to achieve a maximum

ON-current. Fig. 3(a) shows the ON-current ($I_{on}$) versus doping density in MoS$_2$ based JL-FET and Conv. FET for high-performance (HP) (upper plot) and low-standby-power (LSTP) (lower plot) technologies at 5.9 nm node. As schematically shown in Fig. 3(b), $I_{on}$ is obtained by setting OFF-current ($I_{off}$) and supply voltage ($V_{dd}$) to corresponding values listed in Table I. Mobility is set to be 60 cm$^2$/V s, the theoretically predicted upper limit of electron mobility in monolayer MoS$_2$ in high-K gate dielectric environment [13]. It can be observed in Fig. 3(a) that HP prefers higher doping density, while LSTP restricts the doping density to within 1×10$^{20}$ - 3×10$^{20}$ cm$^{-3}$. This can be explained by the different device working regimes of HP and LSTP. As shown in Fig. 3(b), HP mostly works in the super-threshold regime, thus it is more sensitive to the ON-state resistance (source/drain series resistances play the primary role in determining $I_{on}$), and less affected by SS, even in the case of very high doping density. In contrast, LSTP mostly works in sub-threshold regime, thus the merit of lower source/drain series resistance with higher doping density begins to be overwhelmed by degraded SS at very high doping density. On the other hand, JL-FETs show higher $I_{on}$ compared to Conv. FETs in high doping condition, which is due to their better device electrostatics.

In the JL-FET design, high channel doping induces two possible issues. One is that it shifts the threshold voltage toward negative (for n-type device) direction, making the device work in ON-state at zero bias, which is not desirable in low-power applications. To tune the threshold voltage to positive value, gate metals with high work-function should be used. The yellow circle-line in Fig. 3(c) represents required work-function (WF) to tune the threshold voltage to ~0.2V for a given doping density.

The inset in Fig. 3(c) shows the WF range (the height of the blue squares) of some typical metals, which indicates that the threshold voltage of JL-FETs with the studied doping density range can indeed be tuned to be positive by choosing the proper gate metals. This merit stems from the atomic scale thickness of 2D materials that guarantees a reasonably low impurity area density $\rho_{dop}$ at ultrahigh impurity volume density $N_{dop}$, thereby lowering the required WF, i.e., $\rho_{dop} = N_{dop}T_{Ch} \propto \alpha + \varepsilon_{OX}WF/T_{OX}$, where $T_{Ch/OX}$ is the channel/gate dielectric thickness, $\alpha$ is a device geometry dependent factor, and $\varepsilon_{OX}$ is the gate dielectric permittivity. From this perspective, compared to bulk materials based JL-FET, 2D material based JL-FET allows higher volume doping level, which is desired for contact engineering [7],[9]. The other issue is that high doping induces impurity scattering and degrades carrier mobility, which is a commonly observed and well-studied effect in bulk semiconductors. Although this effect in monolayer MoS$_2$, as well as in other 2D semiconductors, has not been quantitatively studied in experiments because of lack of effective doping methods at this stage, it would be insightful to consider a typical condition by adopting a heuristic mobility model widely applied for Si [14],

$$\mu = \mu_{min} + \frac{\mu_0}{1+(N_{dop}/N_{ref})^\beta} \qquad (1)$$

where $\mu_{min}$ is the minimum mobility, $\mu_0$ is the difference between the maximum and minimum mobility expected. $N_{ref}$ is the reference doping density, $\beta$ is a fitting power index. Parameter values are listed in Fig. 3(c). Note that $\mu_{min}$ (corresponding to the highest doping density) is set to be a low value reported in literature [3], which may be partially due to unintentional high doping density. Since we have fixed the minimum and maximum mobility that come from experimental data, it is expected that the error introduced by the estimated $\beta$ and $N_{ref}$ should not be too large to affect the drawn conclusion. Calculated $I_{on}$ considering mobility degradation is shown in Fig. 3(d). For HP application, $I_{on}$ for both Conv. and JL-FETs increase first until doping densities reach ~ 2×10$^{20}$ cm$^{-3}$, and then decrease. The initial increase is due to reduced source/drain series resistance as in Fig. 3(a), while the decrease is due to the effect of mobility degradation becoming dominant. The overall $I_{on}$ of JL-FETs is lower than that of Conv. FET, which is due to the fact that the carriers in JL-FETs suffer from extra impurity scattering in the channel region compared to Conv. FETs. For LSTP application, combined effects of electrostatics and mobility degradation with high doping density screen the merit of reduced source/drain series resistance in the studied doping range, leading to a monotonous decrease of $I_{on}$ with doping density. Note that $I_{on}$ of JL-FET decreases more slowly compared with that of Conv. FET due to its better electrostatics. Based on these observations, it can be concluded that $MoS_2$ JL-FET structure can always be used for LSTP applications because LSTP is not very sensitive to carrier mobility. In contrast, HP applications relies heavily on carrier mobility, thus potential use of JL-FET structure depends on the degree of mobility degradation with doping density in monolayer $MoS_2$, which requires further experimental studies.

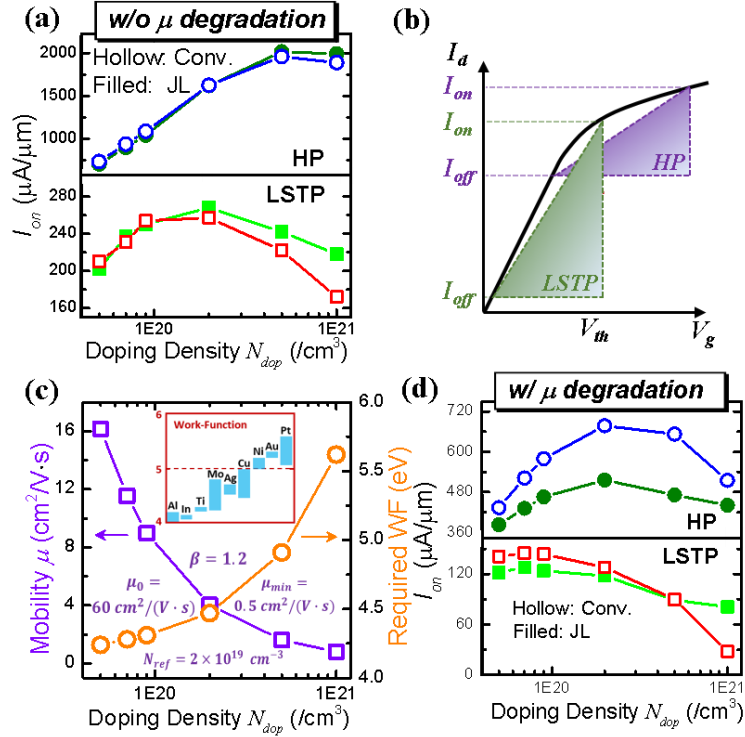

Fig. 3. (a) $I_{on}$ versus doping density $N_{dop}$ for JL and Conv. FET for HP (upper) and LSTP (lower) applications. Mobility degradation with doping density is not considered. (b) Extraction of $I_{on}$ for HP and LSTP. (c) A heuristic mobility model for monolayer $MoS_2$ to take into account the impurity scattering. These values are tuned to be in the experimentally obtained mobility range. (d) The same as (a), but mobility degradation with doping density is considered.

In summary, in order to lower the fabrication cost, and improve the device performance of FETs in sub-10 nm nodes, the combination of JL-FET structure and 2D semiconductor (using $MoS_2$ as an example) is studied through rigorous dissipative quantum transport simulations. It is found that 1) $MoS_2$ based FETs outperform their Si counterpart; 2) the JL-FET structure further improves the device electrostatics and ON-current of $MoS_2$ based FETs; 3) doping density should be optimized to reach the highest ON-current; and 4) the threshold voltage of $MoS_2$ based JL-FETs can always be tuned to be positive by using proper gate metals, which allows a higher volume doping level desired for contact engineering. However, it is also found that possible impurity doping induced mobility degradation in JL-FETs may be an issue that could mask the competence of this structure for high-performance applications of $MoS_2$ based JL-FETs. The bottom line is that JL-FET structure can be used in low-power applications, which do not heavily rely on carrier mobility. It is worthwhile to note that although only $MoS_2$ is studied in this work, the conclusions drawn in this study is applicable to other semiconducting members in the TMD family, because of their essentially similar properties.

**Acknowledgment**

This work was supported by the Air Force Office of Scientific Research (AFOSR), Arlington, VA, USA, under Award FA9550-14-1-0268 (R18641).